\documentclass {ivcnz}

\usepackage{ulem}

\usepackage[english,british]{babel}  
\usepackage[utf8]{inputenc}          
\usepackage{url}                     
\usepackage{xspace}                  
\usepackage{csquotes}                
\usepackage{enumitem}                
\usepackage{calc}                    
\usepackage[printonlyused]{acronym}  
\usepackage{amsmath}

\usepackage{tabularx}  
\usepackage{booktabs}  

\usepackage[pdftex,final]{graphicx}
\DeclareGraphicsExtensions{.pdf,.jpg,.png}
\graphicspath{{./images/},{./images/}{./images/}}

\newlength{\PicSep}
\acrodef{VE}{virtual environment}
\acrodef{HCP}{Head Coupled Perspective}
\acrodef{MOG}{Multiplayer Online Game}

\title{New Zealand involvement in Radio Astronomical VLBI Image Processing}
\author{Stuart D. Weston$^1$, Tim Natusch$^1$, Sergei Gulyaev$^1$}

\address{$^1$Institute for Radio Astronomy and Space Research , AUT University, New Zealand.\\
  Email: \url{sweston@aut.ac.nz}
}

\begin{document}

\begin{abstract}
With the establishment of the AUT University 12m radio telescope at Warkworth, New Zealand has now become a part of the international Very Long Baseline Interferometry (VLBI) community. A major product of VLBI observations are images in the radio domain of astronomical objects such as Active Galactic Nuclei (AGN). Using large geographical separations between radio antennas, very high angular resolution can be achieved. Detailed images can be created using the technique of VLBI Earth Rotation Aperture Synthesis. We review the current process of VLBI radio imaging. In addition we model VLBI configurations using the Warkworth telescope,  AuScope (a new array of three 12m antennas in Australia) and the Australian Square Kilometre Array Pathfinder (ASKAP) array currently under construction in Western Australia, and discuss how the configuration of these arrays affects the quality of images. Recent imaging results that demonstrate the modeled improvements from inclusion of the AUT and first ASKAP telescope in the Australian Long Baseline Array (LBA) are presented. 
\end{abstract}

\keywords{Radio Astronomy, interferometry, VLBI, eVLBI, Image Processing, New Zealand, LBA, ASKAP, AuScope}

\maketitle


\long\def\symbolfootnote[#1]#2{\begingroup%
\def\thefootnote{\fnsymbol{footnote}}\footnote[#1]{#2}\endgroup\vspace{-30pt}}

\section{Introduction}

The Warkworth Radio Astronomical Observatory (WRAO) is located some 60 km north of the city of Auckland, near the township of Warkworth. The observatory is operated by the Institute for Radio Astronomy and Space Research (IRASR) of AUT University. The observatory's 12-m radio telescope operates in three frequency bands centred on 1.4, 2.3 and 8.6 GHz. In addition to astrophysical observations this fast-slewing ( $5\,^{\circ}$  per second in Azimuth) antenna is well suited to the purposes of geodetic VLBI \cite{IVSWebPage} and spacecraft navigation and tracking \cite{NZ12m}.

In February 2011 the AUT University 12m radio antenna officially joined the Australian Long Baseline Array (LBA) and now regularly participates in its VLBI sessions. The primary product of this VLBI work is high resolution radio domain images from which the physical properties of radio astronomical sources are studied. 

The AUT University 12m radio antenna has expanded the maximum east-west baselines of the LBA \cite{LBA} by almost a factor of two, from $\approx 1300$km to $\approx 2400$km. With the further addition of the ASKAP antennas in Western Australia east-west baselines of $\approx 5000$ km are achieved, providing corresponding increases in resolution. The goal of this paper is to investigate the way for further improvement of the quality of the array, first of all in terms of extension of the north-south baselines to yet further improve the resultant image quality.

In Section 2, we briefly present the theory behind radio interferometry and review the current methods used to obtain radio images of astronomical sources. Section 3 specifically outlines current image recovery methods. In Section 4, we model the effects of adding the AUT University 12m radio antenna which will be referred too as Warkworth, the new ASKAP \cite{ASKAP} and AuScope \cite{AuScope} antennas on the imaging performance of the LBA. Section 5 discusses current activities and presents actual images achieved with the AUT 12m antenna as part of the LBA.


\section{Radio interferometry}

A radio interferometer is a pair (or more) of antennas used to measure the visibility function due to the sky brightness within the field of view of the antennas \cite{Thompson2004}. From sampling this visibility function it is possible to recover an image of the observed field of view \cite{Thompson2004}.

\subsection{UV coverage for models}

Imaging of a radio astronomical source by the technique of aperture synthesis was first demonstrated by Prof. Ryle \cite{Ryle1960} using the Cambridge Radio Telescope \cite{Ryle1962}. Antennas able to track a source for an extended period as the Earth rotates will trace out elliptical paths in the $u,v$ plane (orthogonal plane to the direction of the astronomical source). The components $u,v,w$ may be determined from the expression  \cite{Thompson2004}:

$ \left( \begin{smallmatrix} u\\v\\w \end{smallmatrix} \right) = \frac{1}{\lambda} \left( \begin{smallmatrix} sinH_0&cosH_0&0\\-sin\delta_0 cosH_0 & sin\delta_0 sinH_0 & cos \delta_0\\cos\delta_0 cosH_o&-cos\delta_0sinH_0&sin\delta_0 \end{smallmatrix} \right) \left( \begin{smallmatrix} L_x\\L_y\\L_z \end{smallmatrix} \right)$

Here $H,\delta_0$ are the hour angle and declination of the source, $\lambda$ is the wavelength of the radio frequency being observed. It is customary in VLBI to eliminate the $H_0$ (hour angle) term by setting the x axis of the coordinate system to the Greenwich meridian $H_0=0$, resulting in:

\begin{equation} \label{eq:uvelipse}
          u^2 +  \left( \frac {v -(L_z/\lambda)cos\delta_o}{sin\delta_o} \right) ^2 = \frac{ L_x^2 +L_y^2}  {\lambda}     
\end{equation}

As $L_x$,$L_y$ and $L_z$ are constants for a given pair of antennas, this is the equation of an ellipse in the u,v plane (it becomes the equation of a circle $u^2+v^2$ when $\delta_0 = 90\,^{\circ}$). For an array of $N$ antennas we will have $N(N-1)/2$ pairs of elliptical loci.

\begin{figure}[h]
	\centering%
	\includegraphics[width=\columnwidth]{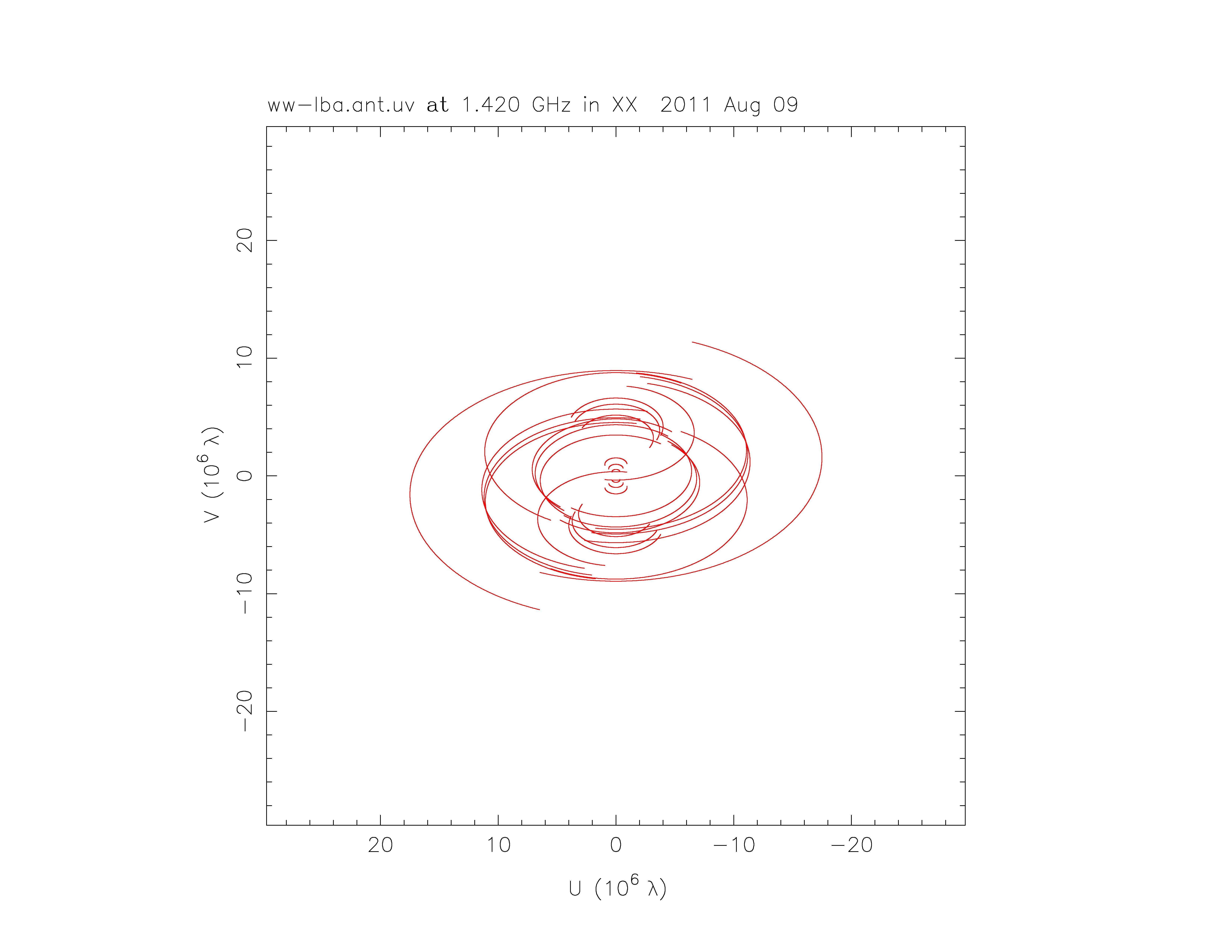}
	\caption{UV Coverage for Warkworth and the LBA}
	\label{fig:wwlbauv}
\end{figure}

\begin{figure}[h]
	\centering%
	\includegraphics[width=\columnwidth]{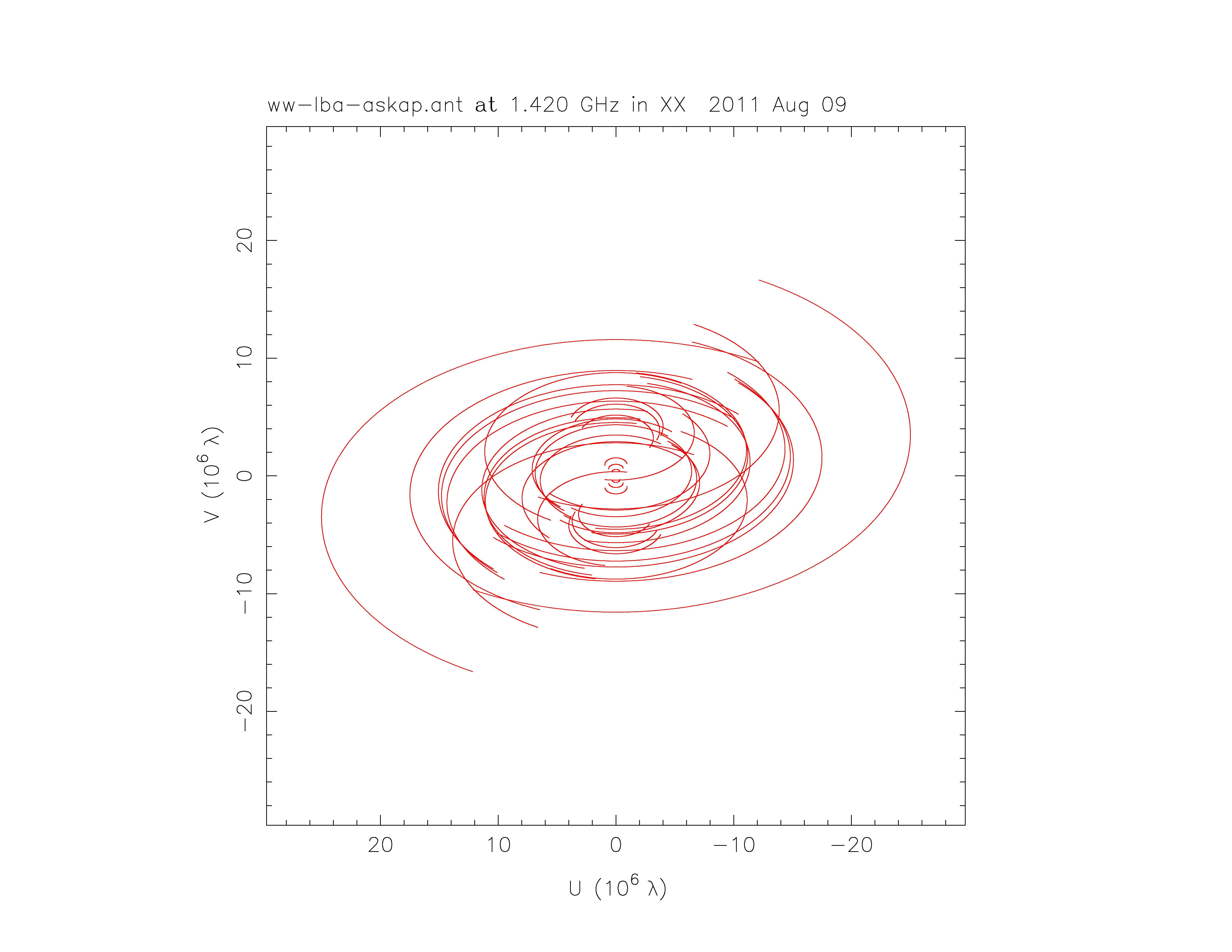}
	\caption{UV Coverage for Warkworth, LBA and ASKAP}
	\label{fig:wwlbaaskapuv}
\end{figure}

\begin{figure}[h]
	\centering%
	\includegraphics[width=\columnwidth]{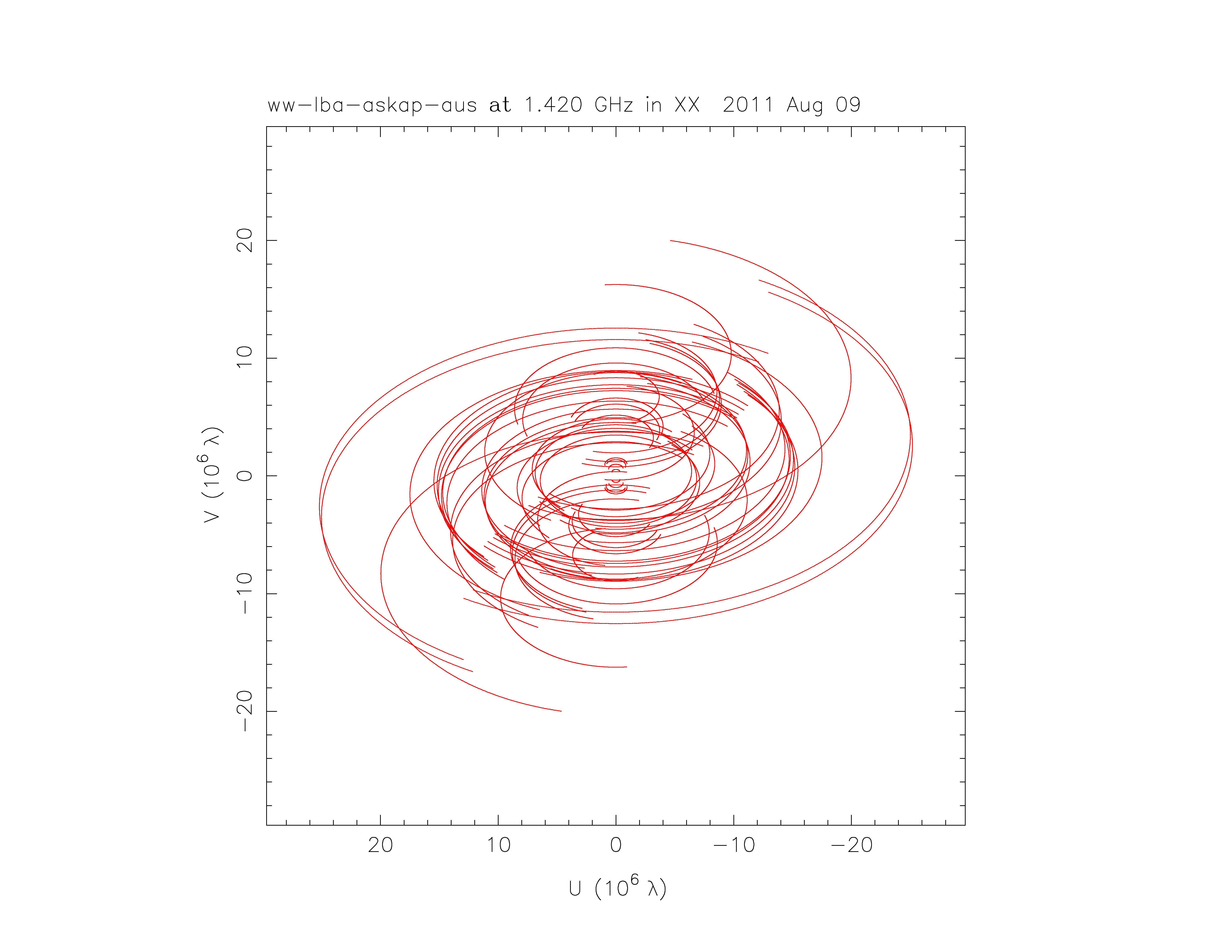}
	\caption{UV Coverage for Warkworth, LBA, ASKAP and AuScope}
	\label{fig:wwlbaaskapauscopeuv}
\end{figure}

Plots of these loci (tracks) demonstrate the progressive improvement in filling of the $u,v$ plane as additional antennas are added to the LBA array: Warkworth plus LBA in Figure~\ref{fig:wwlbauv}, Warkworth, LBA and ASKAP in Figure~\ref{fig:wwlbaaskapuv}, Warkworth, LBA, ASKAP and AuScope in Figure~\ref{fig:wwlbaaskapauscopeuv}. All plots are for a common source declination of $-37\,^{\circ}$, that is for a Southern Celestial Hemisphere radio source. Both $u$ and $v$ are given in wavelengths, $\lambda$.

\subsection{Visibility function}

\begin{figure}[!h]
	\centering%
	\includegraphics[width=\columnwidth]{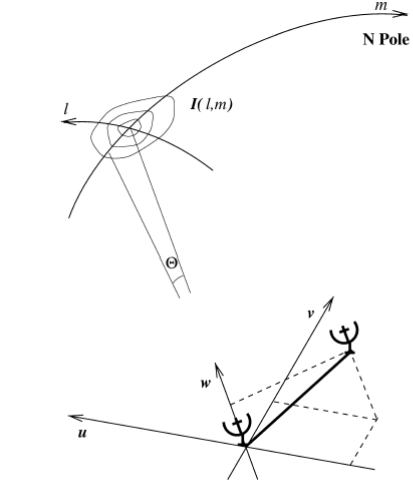}
	\caption{Geometric relationship between the source observed $I(l,m)$ and an interferometer or antenna pair upon the $u,v$ plane \cite{Diamond2003}}
	\label{fig:uvplane}
\end{figure}

The astronomical source is treated as a two dimensional image of intensity $I(l,m)$ on the celestial sphere, $l$ and $m$ being coordinates in the plane of the celestial sphere. The projection of $l$ and $m$ to the plane perpendicular to the direction of the astronomical source from the Earth, as defined by the coordinate system $u$ and $v$, is shown in Figure ~\ref{fig:uvplane}.

It can thus be seen that the visibility measured by an interferometer is a sample of the visibility function $V(u,v)$, and may be expressed as the Fourier transform of the modified sky intensity $I$  \cite{Thompson2004}:

\begin{equation} \label{eq:visibility}
     V(u,v)=\int \int I(l,m)e^{-2\pi j(ul+vm)}dl dm
\end{equation}

As the visibility is sampled out to a maximum radius $b_{max}$ in the $u,v$ plane, the array produces information similar to a single circular aperture of diameter $D \approx \lambda b_{max}$. Resolution of the image will be $ \approx 1/b_{max}$, but the quality of the image is determined by the $u,v$ coverage.


\section{Image recovery methods}
\label{s:Methods}

During an observation, sources are tracked by the interferometer array and the signal at each telescope separately recorded. Data from all telescopes is then sent to a processing centre for correlation. As $V$ is a Fourier transform of the source brightness distribution $I$ Equation (\ref{eq:visibility}), the latter can be recovered by means of the inverse Fourier transformation:

\begin{equation} \label{eq:SkyBrightness}
     I(l,m)=\int \int V(u,v)e^{2\pi i(ul+vm)}du dv
\end{equation}

As integration occurs over $u$ and $v$, the more information we have about $V(u,v)$, the more fully we can conduct numerical integration in \ref{eq:SkyBrightness}. In other words: the better the $u,v$ coverage, the greater chance to restore the image, $I(lm)$, without errors and artefacts.

Prior to the inverse Fourier transforming \ref{eq:SkyBrightness}, calibration steps are invariably required as variations in performance and characteristics of antennas must be accounted for (the array is not homogeneous). Additionally, data may have to be flagged due to factors such as equipment failure at an antenna or occurrence of radio frequency interference (RFI).

To take advantage of the computational efficiency of the Fast Fourier Transform (FFT), the correlated visibilities must be binned onto a regular grid. Weighting schemes may be applied at this point that influence this gridding and aspects of the fidelity of the final image \cite{CornwellBraunBriggs1999}. Once ``gridded'' the data is Fourier transformed to obtain a first cut image. This first image is commonly referred to in radio astronomy as the ``dirty map'' as it consists of the true image of the source convolved with (corrupted by) the point spread function (PSF) of the imaging instrument, the VLBI array itself. Through use of one of various algorithms noted in the following subsection, an approximation to the true image is sought through deconvolution of this dirty map. 

\subsection{Algorithms}

It was shown in Section 2 that the u,v plane is incompletely sampled by the antennas, thus an analytic approach cannot be used in recovering the true image. Several numerical algorithms have been found to effect a solution and are commonly used in radio astronomy interferometry to recover an image; CLEAN, MEM, NNLS. We only briefly present these different algorithms (more complete explanation and comparison is available from Cornwell \cite{CornwellBraunBriggs1999}) with more detail about CLEAN the most commonly used in VLBI imaging.

{\bf CLEAN} The most widely used radio astronomical image deconvolution technique is the CLEAN algorithm. Devised by H\"{\o}gbom \cite{Hogbom1974}, this is a numerical deconvolution process applied in the $(l,m)$ domain. The basic assumption is that the true image may be modeled as a set of point sources. In an iterative process the amplitudes and positions of these point sources are established through fitting the PSF of the VLBI array to the brightest point on the image. A final image is produced by convolution of this point source model with an appropriately sized Gaussian restoring PSF. There have been more recent refinements to this algorithm such as multi-scale CLEAN \cite{Cornwell2008}.

{\bf MEM} The maximum entropy deconvolution algorithm \cite{Cornwell1985}.

{\bf NNLS} Non-Negative Least Squares (NNLS) is discussed in some detail by Briggs \cite{Briggs1995}.

\subsection{Future trends}

{\bf Compressive sampling} Until recently it has been held that to recover a signal or image the sampling rate must be at least twice the maximum frequency present in the signal \cite{Shannon1949}, at or above the Nyquist rate. A result of the newly developing field of compressive sampling finds that signals sampled at less than the Nyquist rate can be recovered if the information is ``sparse''.  It was shown in Section 2 that the $u,v$ plane is not completely sampled by the antennas and thus we have a sparse model. Recently, some work has been undertaken to see if Compressive sampling can be used to recover images in the radio domain \cite{FengLi2011}. Feng Li et al  \cite{FengLi2011} are attempting to implement these methods for deconvolving images in semi-real time. If this could be conducted in sequence with eVLBI correlation, then for the first time images could be available in near real time after an observation.

\subsection{Packages}

There are several software packages widely used by the Radio Astronomy community incorporating one or more of the above algorithms. These are the Astronomical Image Processing System (AIPS) \cite{AIPS}, MIRIAD \cite{MIRIAD} and the standalone package DIFMAP for producing images from radio interferometers using the CLEAN deconvolution \cite{Shepherd1995} algorithm.

For the next generation of radio astronomical telescopes the Common Astronomy Software Applications (CASA)\cite{CASA} is being developed by an international consortium, built on C++ tools under an iPython interface. This includes in addition to CLEAN and MEM a module for deconvolution using NNLS.


\section{Modeling NZ involvement in VLBI}
\label{s:Modeling}

\subsection{Modeling}

\begin{figure}[!h]
	\centering%
	\includegraphics[width=\columnwidth]{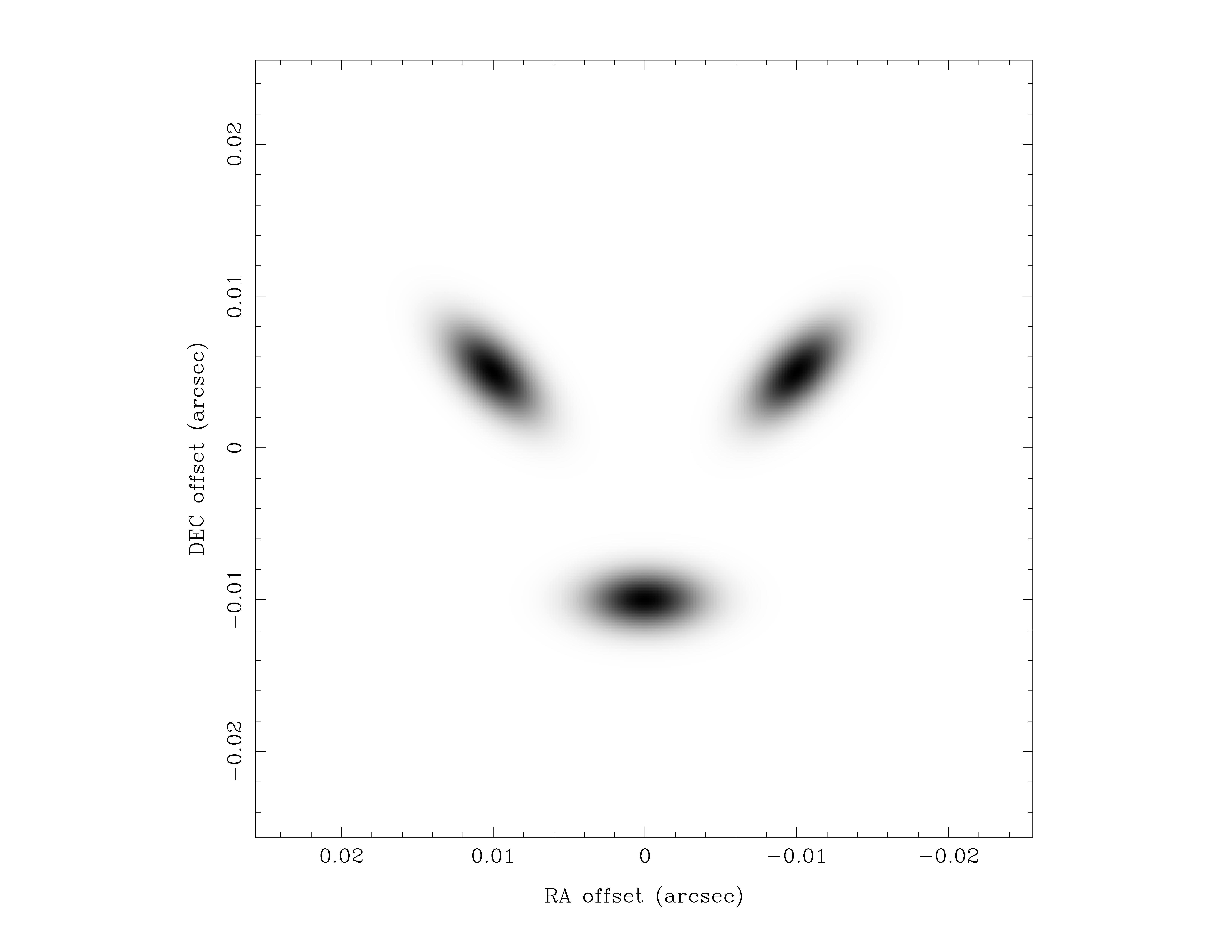}
	\caption{Galaxy triplet model image}
	\label{fig:glxytriplet}
\end{figure}

Consider the model image in Figure~\ref{fig:glxytriplet} of three elliptical Gaussian sources to represent a radio galaxy triplet.

\begin{figure}[h]
	\centering%
	\includegraphics[width=\columnwidth]{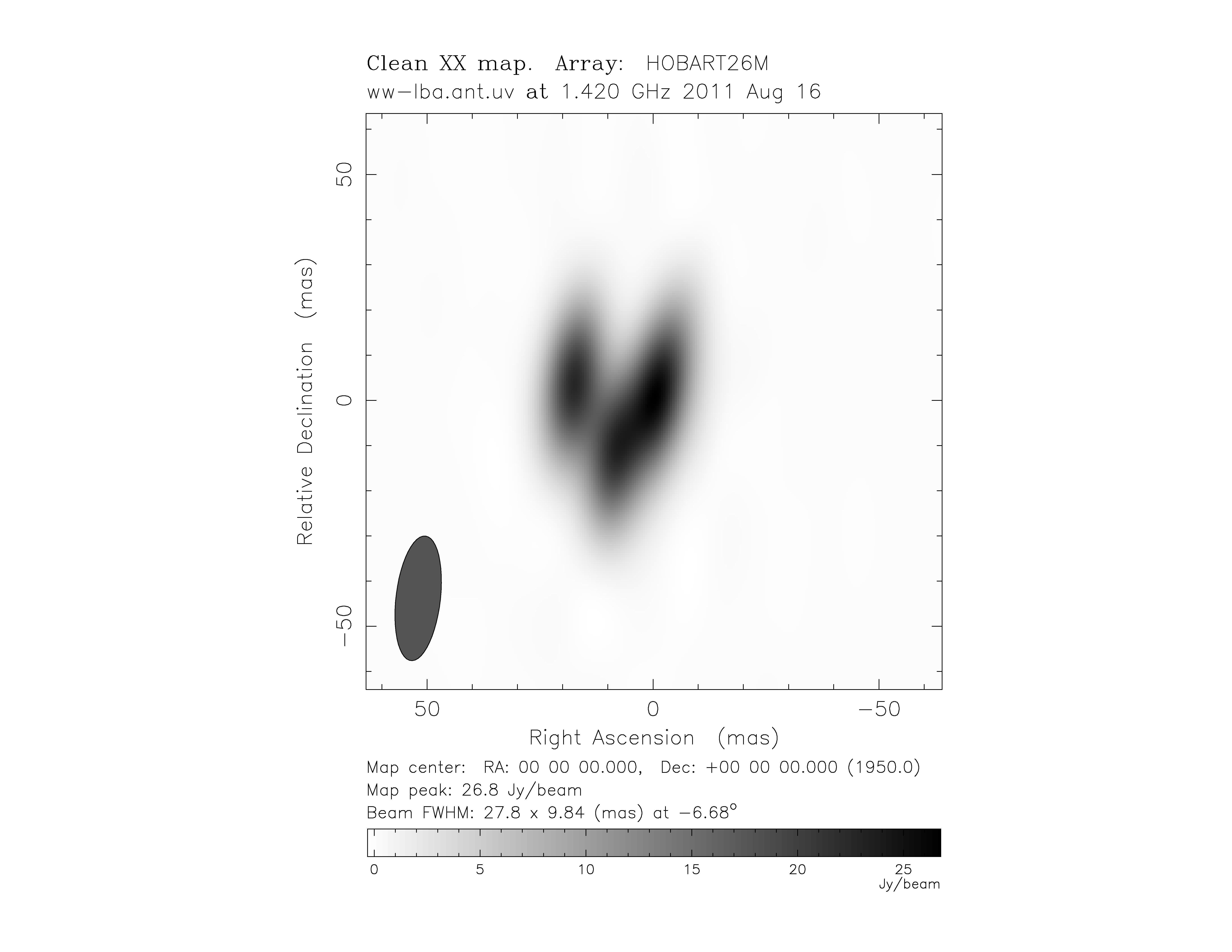}
	\caption{CLEAN Image from Warkworth LBA}
	\label{fig:clean_ww-lba}
\end{figure}

This model image was input to a process developed at AUT\cite{Weston2008} that simulates and models VLBI arrays using a combination of the packages MIRIAD and DIFMAP. First taking the array configuration of Warkworth and the LBA we obtain the image presented in Figure~\ref{fig:clean_ww-lba}. Without prior knowledge of the model image we would most likely conclude that we had a double source with one large extended component to the right, a second more compact component to the left. Note that the restoring beam shape (bottom left hand corner solid shaded ellipse) influences the final image obtained. 

\begin{figure}[h]
	\centering%
	\includegraphics[width=\columnwidth]{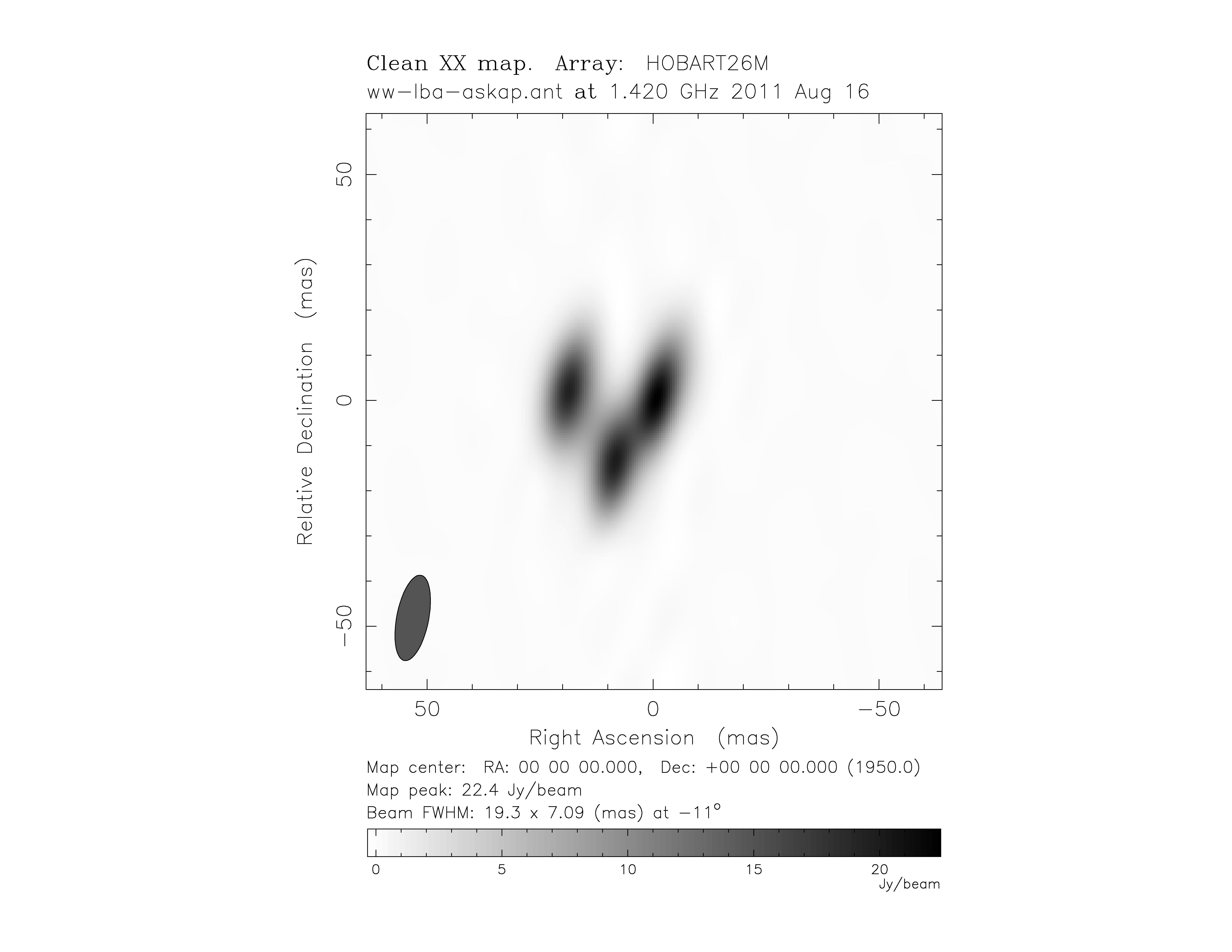}
	\caption{CLEAN Image from Warkworth, LBA and ASKAP}
	\label{fig:clean_ww-lba-askap}
\end{figure}

Following inclusion of an ASKAP antenna into the simulation as shown in Figure~\ref{fig:clean_ww-lba-askap} it is just possible to resolve three separate components of the source, although the separation between two of the components is minimal. The beam (bottom left corner of image) is clearly more compact as a result of the additional resolution available. 

\begin{figure}[h]
	\centering%
	\includegraphics[width=\columnwidth]{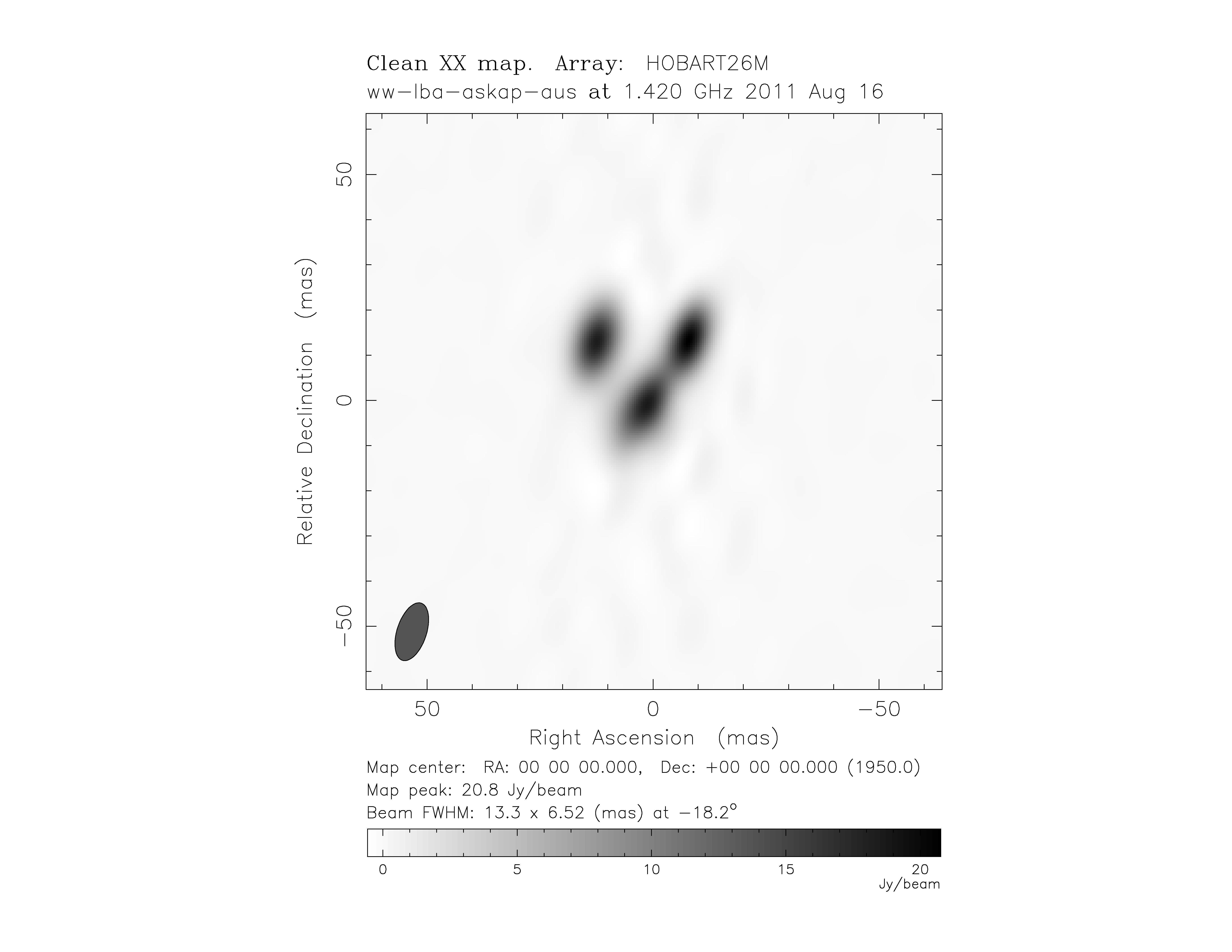}
	\caption{CLEAN Image from Warkworth, LBA, ASKAP and AuScope}
	\label{fig:clean_ww-lba-askap-auscope}
\end{figure}

Finally, inclusion of the AuScope antennas adds additional and longer n-s baselines. A significantly more compact, although not quite circular beam results and three unambiguously separate sources are resolved as shown in Figure~\ref{fig:clean_ww-lba-askap-auscope}. 

\pagebreak[4]

\section{Recent results}

\subsection{VLBI}

\begin{figure}[!h]
	\centering%
	\includegraphics[width=\columnwidth]{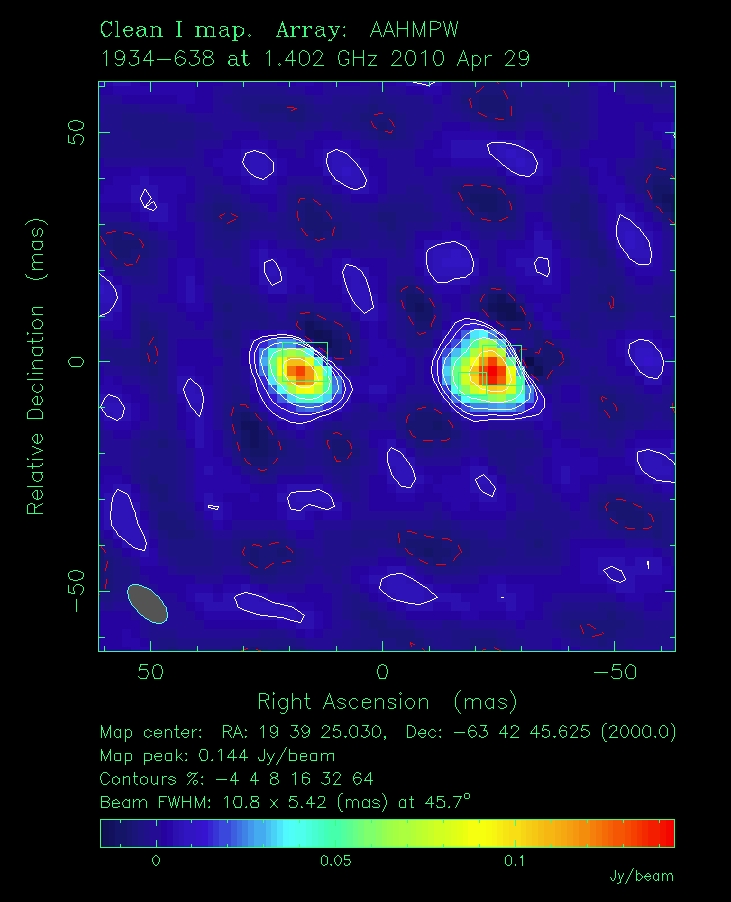}
	\caption{VLBI Image of 1934-638 at 1.4 GHz using LBA, Warkworth and ASKAP, Credit: S. Tingay}
	\label{fig:VLBIImage01}
\end{figure}

\begin{figure}[!h]
	\centering%
	\includegraphics[width=\columnwidth]{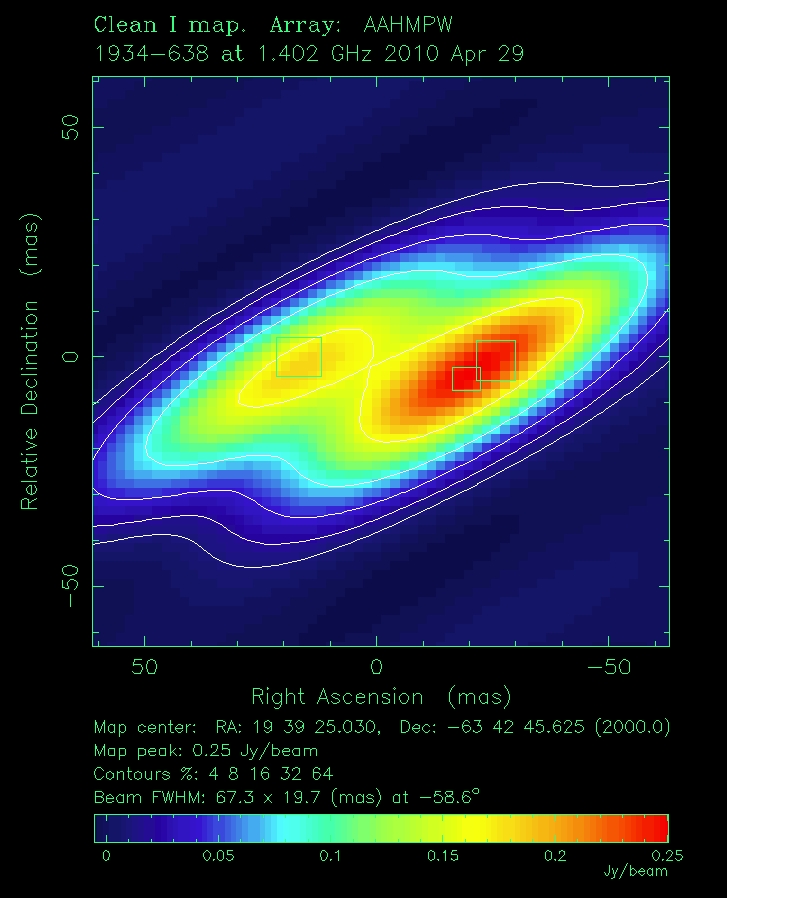}
	\caption{VLBI Image of 1934-638 at 1.4 GHz by the LBA, Credit: S. Tingay}
	\label{fig:VLBIImage02}
\end{figure}

The AUT University 12m radio antenna had been involved with the LBA prior to 2011 on an ad-hoc basis, but is now routinely included in experiments. First LBA science results that include the AUT and ASKAP antennas were achieved on April 2010 with images of PKS 1934-638 obtained at 1.4 GHz \cite{Tzioumis2010} as shown in Figure~\ref{fig:VLBIImage01}. For comparison to demonstrate the improvement in imaging achieved compare this with an image taken of the same source using the LBA configuration before the addition of Warkworth and ASKAP Figure~\ref{fig:VLBIImage02}.

\begin{figure}[!h]
	\centering%
	\includegraphics[width=\columnwidth]{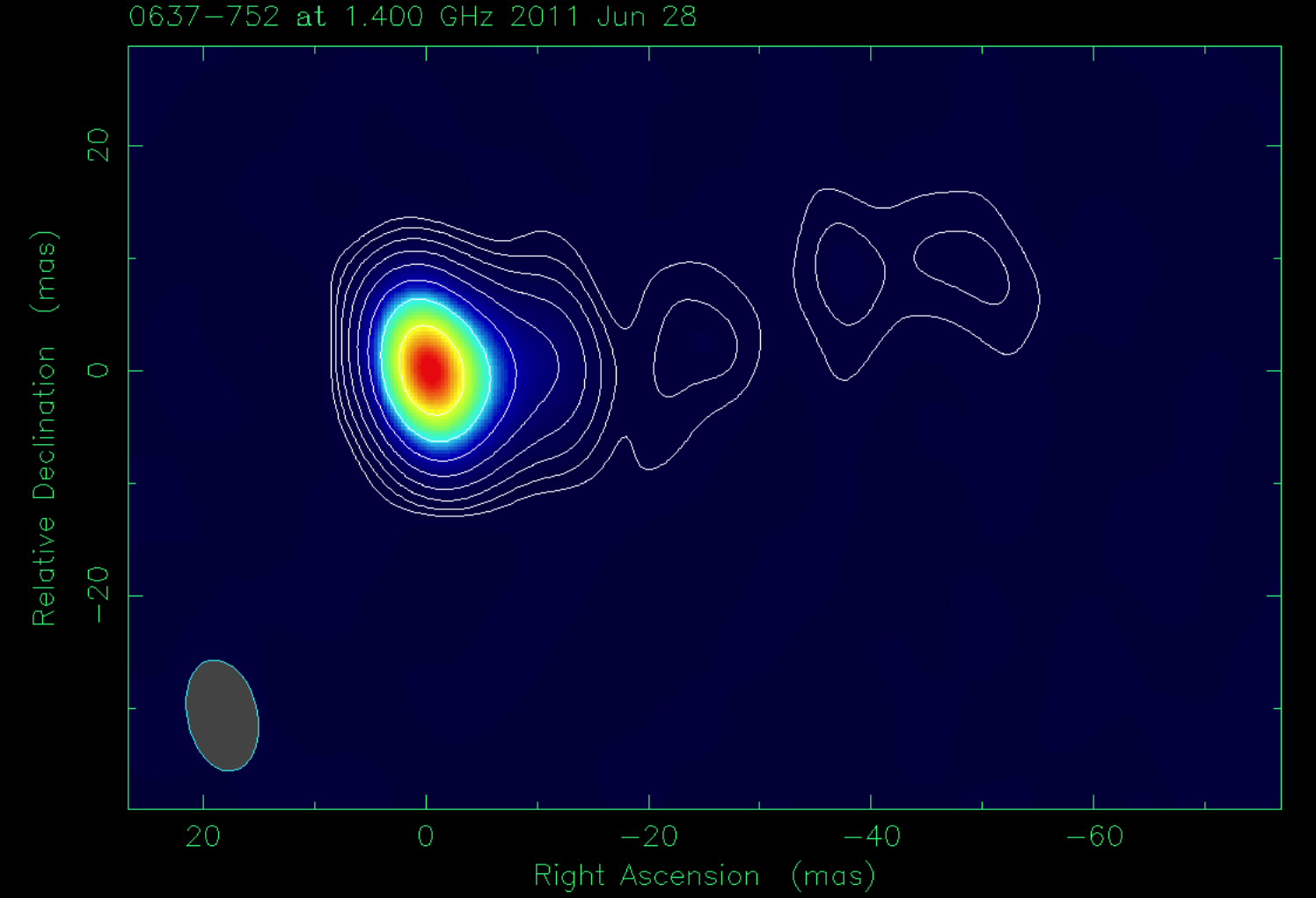}
	\caption{eVLBI Image of 0637-752 at 1.4 GHz by Warkworth, LBA and ASKAP July 2011, Credit: S. Tingay}
	\label{fig:eVLBIImage}
\end{figure}

\subsection{eVLBI}

eVLBI involves sending the received data from all antennas in real-time over a high speed network to the correlator. In July 2011 the first such LBA observations that included the AUT 12m antenna and newly commissioned ASKAP antenna took place with a source observed in L-Band (1.4 GHz), with the resultant image shown in Figure~\ref{fig:eVLBIImage}. The data from the AUT University 12m radio antenna was streamed across the Kiwi Advanced Research and Education Network (KAREN) in real time at a rate of 512 Mbps to the correlation centre at Curtin University, WA. 

\pagebreak[4]


\begin{thebibliography}{21}

\bibitem{IVSWebPage} IVS Home Page (n.d), Available: International VLBI Service for Geodesy and Astrometry Web site: http://ivscc.gsfc.nasa.gov/

\bibitem{NZ12m} S. Gulyaev and T. Natusch, New Zealand 12-m VLBI Station, International VLBI Service for Geodesy and Astrometry 2009 Annual Report, NASA/TP 2010-215860. Eds.: D. Behrend and K. D. Baver, 2010, pp. 138-141.

\bibitem{LBA} The Australian Long Baseline Array (n.d), CSIRO, Available: http://www.atnf.csiro.au/vlbi/

\bibitem{ASKAP} Australian Square Kilometre Array Pathfinder Home Page (n.d), CSIRO, Available: http://http://www.atnf.csiro.au/SKA/

\bibitem{AuScope} AuScope Home Page (n.d), An Organisation for National Earth Science Infrastructure Program, Available: http://www.auscope.org/index.php

\bibitem{Thompson2004} Thompson, A. R., J. M. Moran, G.W. Swenson . (2004). Interferometry and Synthesis in Radio Astronomy, WILEY-VCH.

\bibitem{Ryle1960} Ryle, M. and A. Hewish (1960). The synthesis of large radio telescopes. Monthly Notices of the Royal Astronomical Society 120: 220. 

\bibitem{Ryle1962} Ryle, P. M. (1962). The New Cambridge Radio Telescope. NATURE 194: 517-518.

\bibitem{Diamond2003} Diamond, P. J., S. T. Garrington, et al. (2003). MERLIN User Guide.

\bibitem{CornwellBraunBriggs1999} Cornwell, T.,Braun, R. Briggs D. S., 1999. Synthesis Imaging in Radio Astronomy II ed  G. B. Taylor, C. L. Carilli and R. A. Perley, ASP Conference Series, 180, 1999, 151-170.

\bibitem{Hogbom1974} H\"{\o}gbom, J. A. (1974). Aperture Synthesis with a Non-Regular Distribution of Interferometer Baselines. Astron. Astrophys. Suppl. 15: 417-426.

\bibitem{Cornwell2008}Cornwell, T. J.. Multiscale CLEAN Deconvolution of Radio Synthesis Images, IEEE Jpurnal of Selected Topics in Signal Processing,Vol 2, No 5, October 2008.

\bibitem{Cornwell1985}Cornwell, T. J.,Evans, K. F. (1985). A simple maximum entropy deconvolution algorithm. Astron. Astrophys 143, 77-83.

\bibitem{Briggs1995} Briggs, D. (1995). High Fidelity Deconvolution of Moderately Resolved Sources. Socorro, New Mexico, The New Mexico Institute of Mining and Technology Doctor of Philosophy in Physics.

\bibitem{Shannon1949} C. E. Shannon, Communication in the presence of noise, Proc. Institute of Radio Engineers, vol. 37, no.1, pp. 10–21, Jan. 1949. Reprint as classic paper in: Proc. IEEE, Vol. 86, No. 2, (Feb 1998)

\bibitem{FengLi2011} Li F., Cornwell, T., de Hoog, Frank. The application of compressive sampling to radio astronomy I: Deconvolution.	arXiv:1106.1711v1 [astro-ph.IM]

\bibitem{AIPS} AIPS Home Page (n.d) [Online]. Available: http://www.aips.nrao.edu/

\bibitem{MIRIAD} A retrospective view of Miriad, by Sault R.J., Teuben P.J., Wright M.C.H., 1995. In Astronomical Data Analysis Software and Systems IV, ed. R. Shaw, H.E. Payne, J.J.E. Hayes, ASP Conference Series, 77, 433-436.

\bibitem{Shepherd1995} DIFMAP: an interactive program for synthesis imaging by M. C. Shepherd, T. J. Pearson, and G. B. Taylor [Bull. Amer. Astron. Soc., 26, 987-989 (1994), Bull. Amer. Astron. Soc., 27, 903 (1995).

\bibitem{CASA} CASA Home Page (n.d) [Online]. Available: http://http://casa.nrao.edu/

\bibitem{Weston2008} Weston, S. Development of Very Long Baseline Interferometry (VLBI) techniques in New Zealand: Array simulation, image synthesis and analysis, AUT University Master of Philosophy,  http://hdl.handle.net/10292/449

\bibitem{Tzioumis2010} A. K. Tzioumis, S. J. Tingay, B. Stansby, J. E. Reynolds, C. J. Phillips, S. W. Amy, P. G. Edwards, M. A. Bowen, M. R. Leach, M. J. Kesteven, Y. Chung, J. Stevens, A. R. Forsyth, S. Gulyaev, T. Natusch, J. P. Macquart, C. Reynolds, R. B. Wayth, H. E. Bignall, A. Hotan, C. Hotan, L. Godfrey, S. Ellingsen, J. Dickey, J. Blanchard, and J. E. J. Lovell, “Evolution of the Parsec-Scale Structure of PKS 1934--638 Revisited: First Science with the ASKAP and New Zealand Telescopes”, The Astronomical Journal, 140, November 2010, pp. 1506–1510.


\end{thebibliography}
\end{document}